\documentstyle[eqsecnum,aps,prb]{revtex}

\draft

\begin{document}

\title
{
Three-body correlations in the Nagaoka state\\
on the square lattice 
}

\author{Jun-ichi Igarashi, Manabu Takahashi, and Tatsuya Nagao}

\address{Faculty of Engineering, Gunma University, Kiryu,
         Gunma 376-8515,
Japan}


\maketitle


\begin{abstract}

A three-body scattering theory previously proposed by one of 
the present authors is developed to be applied to the
saturated ferromagnetic state in the two-dimensional
Hubbard model.
The single-particle Green's function is calculated by taking
account of the multiple scattering between two electrons and
one hole. Several limiting cases are discussed and 
the relation to the variational principle is examined.
The importance of the three-body correlation is demonstrated
in comparison with the results of the ladder approximation.
A possible phase boundary for the Nagaoka ground state is presented
for the square lattice, which improves
the previous variational results. 

\end{abstract}


\section{Introduction}

Electron correlations in narrow bands have attracted much attention
after the discovery of high T$_c$ cuprate superconductors.
Since a thorough study of electron correlations is quite complicated 
in actual materials, we often use a simplified model to understand
the essence of the effects.
One of the most widely used models is the Hubbard model,
\cite{hubbard,kanamori,gutzwiller} whose Hamiltonian is given by
\begin{equation}
H=-t\sum_{\langle i,j\rangle\sigma}a_{i\sigma}^\dagger a_{j\sigma} 
 + {\rm H.c.}
  + U\sum_i a_{i\uparrow}^\dagger a_{i\downarrow}^\dagger
 a_{i\downarrow} a_{i\uparrow},
\end{equation}
where $t>0$ and $\langle i,j\rangle$ stands for the nearest neighbor pairs.
Operator $a_{i\sigma}$ represents the annihilation of an electron with
spin $\sigma$ at site $i$.

In the low density of occupied electrons, the correlation effects on this
model are exactly taken into account by considering the multiple scattering 
between two electrons. Using this low-density theory, Kanamori\cite{kanamori}
studied the ferromagnetism of nickel metal (here holes are in low density).
He found that the effective interaction between electrons
is considerably reduced from the bare value $U$ due to the correlated
motion avoiding each other, and that 
the Stoner condition for ferromagnetism is strongly modified.
In order to extend his theory to higher density, we need to include
also the effects of the particle-hole multiple scattering,
which describes the magnon excitation on the ferromagnetic
ground state, and those of the electron-magnon interaction.
Hertz and Edwards\cite{hertz} recognized the importance of the vertex
correction, and determined the electron-magnon vertex by enforcing 
the Ward-Takahashi identity.\cite{matsumoto} However, it is impossible 
to determine the vertex 
on all points in the momentum-frequency space by the identity alone,
and also it is unclear how to include the particle-particle
multiple-scattering effect, which becomes important in the
low density of electrons.
One of the present author has proposed a theory\cite{iga1,iga2} which 
takes account of the multiple scattering between two electrons and
one hole with the use of the Faddeev equation of the three-body 
problem.\cite{iga1,iga2,rucken,gor,hsu}
This theory naturally takes account of the particle-particle 
and particle-hole multiple-scattering channels on equal footing, and thereby
it contains the effect interpreted as the vertex correction
for the electron-magnon interaction.

The three-body scattering theory has a clear meaning of the three-body 
problem on the saturated ferromagnetic ground state, although it is
applicable to the paramagnetic phase by deriving a similar integral equation 
to the Faddeev equation.
In this paper, in order to make clear the physical meaning and 
the reliability of the approximation, we confine ourselves to 
the application to the two-dimensional Hubbard model 
in the context of the instability of the saturated ferromagnetic
ground state. 
It has a nice property interpolating from the low-density limit to the
half-filling limit, and satisfies the exact atomic limit\cite{hubbard} 
as previously proved.\cite{iga1,rucken}
One of the present authors\cite{iga2} has already applied the three-body
scattering theory to the one-dimensional Hubbard model 
in comparison with the exact solution, having calculated the spectral
function for the single-particle Green's function.
Later Ruckenstein and Schmitt-Rink\cite{rucken} have examined the same 
approach in comparison with other theories.
However, the actual calculation in two dimensions has not been carried out yet.
In this paper, we calculate the single-particle Green's function in two
dimensions and determine the quasi-particle energy.
We demonstrate the importance of the three-body correlation
by comparison with the ladder approximation.
We find that a low-energy scale for the quasi-particle energy,
which is much smaller than its value of the ladder approximation,
and that the mass enhancement factor is as large as $\sim 10$
in the strong coupling regime.
We discuss several limiting cases for clarifying the approximation made.
We also show that the present theory satisfies the variational principle.
\cite{gor}
This observation helps us to compare the present theory to other
variational theories. 

As regards the stability of saturated ferromagnetism, 
Nagaoka considered in his pioneering study the case of one hole
in an otherwise half-filled band with $U=\infty$.\cite{nagaoka,thouless}
He proved that the saturated ferromagnetism is realized as ground
state on appropriate lattices due to gaining the kinetic enegy of the hole.
For two holes on a square lattice with $U=\infty$, 
Fang {\em et al.}\cite{fang} found that the ground state is a singlet 
by exact diagonalizations for small clusters.
Doucot and Wen\cite{doucot} found that a long-wave length twisted spin state 
has lower energy than the saturated ferromagnetic state.
On the other hand, it has been noticed that finite-size effects are crucial
and that closed-shell configurations (numbers of holes are 1, 5, 9, $\cdots$)
favor ferromagnetism whereas open-shell configurations (numbers of holes
are 2, 3, 4, 6, $\cdots$) tend to destabilize it.\cite{barbieri,linden}
Quantum Monte Carlo results suggest no ferromagnetic phase for
intermediate $U$,\cite{barbieri,hirsch} but the cluster sizes used 
and the simulation temperatures are rather high.
Therefore, for a thermodynamic concentration of holes, such considerations
for a few holes do not serve as proper guides.

Another route to investigate the stability of the Nagaoka 
ground state is a variational study for the excitation energy
with respect to overturning
an up-spin electron and placing it on a down-spin band.
Of course this approach gives only an upper limit for the ferromagnetic
boundary. Roth\cite{roth} was the first to discuss in detail the instability of
the Nagaoka ground state with an approximate scheme
equivalent to using a variational wave function. It is found superior 
to the Gutzwiller function,
since it has more variational flexibility.\cite{allan} 
The similar variational wave function was 
employed by Shastry, Krishnamurthy, and Anderson,\cite{shastry}
for discussing the instability of the Nagaoka ground state
and the spin-wave stiffness. It was also used to calculate
the spin-wave spectrum with general momentum in the $U=\infty$ limit
by Basile and Elser\cite{basile}, and for finite $U$ by Okabe.\cite{okabe}
The softening of the spin-wave energy may set another condition
for the instability of the Nagaoka state.
These calculations leads to an instability for a finite concentration
of holes even in the $U=\infty$ limit. 
Another interesting variational wave function is constructed by
Richmond and Rickayzen,\cite{richmond} who froze
the motion of the down-spin electron and solved exactly the states
for up-spin electrons under the static potential due to the down-spin
electron (see Appendix B).
This study merely leads to the conclusion that the Nagaoka state 
is always stable in the $U=\infty$ limit for any concentrations of holes.
Recently, Hanish, Uhlig, and M\"uller-Hartman\cite{hanish} analyzed
Roth's wave function and its slight extension on various lattices,
and calculated the phase diagram for the Nagaoka ground state.
The present theory is shown to be equivalent to using
a variational wave function similar to Roth's but assuming
the most general form with the states containing
one particle-hole pair. It is more flexible than that 
of Hanish {\em et al.},\cite{hanish} and thereby we obtain the phase diagram 
with a smaller area of the Nagaoka ground state than Hanish {\em et al.}
This may serve as a measure of the accuracy of the present theory.

In Sec.~II, the three-body scattering theory is formulated 
for the single-particle Green's function, and several limiting formulas are 
derived. The relation to the variational method is also discussed. 
In Sec.~III, the numerical solution is presented for the square lattice.
Section IV contains concluding remarks.
In the Appendix A, the explicit solution of the Faddeev equation
is derived in the $U=\infty$ limit, and in the Appendix B
the method of Richmond and Rickayzen is adapted for numerical computations.

\section{Three-Body Scattering Theory}

We briefly summarize the three-body scattering theory,
and discuss several limiting formulas, which are useful to make clear
the physical meaning of the approximation made.
We also discuss the variational character of the theory.

\subsection{The Green's function formalism}

We start by assuming the saturated ferromagnetic
ground state, where only up-spin states are occupied by electrons.
No interaction is working between electrons with the same spin. 
Placing a down-spin electron into the system, we study its motion
using the single-particle Green's function.

We make a perturbational expansion, where the Hartree-Fock energy 
is included into the unperturbed part. Thus the Hamiltonian is divided
into two parts: 
\begin{equation}
 H=H_0+H_1,
\end{equation}
where
\begin{eqnarray}
 H_0 & = & \sum_{\bf k}[\epsilon_{\bf k}
      a_{{\bf k}\uparrow}^\dagger a_{{\bf k}\uparrow}
    +(\epsilon_{\bf k}+Un)
      a_{{\bf k}\downarrow}^\dagger a_{{\bf k}\downarrow}] ,\\
 H_1 & = & \frac{U}{N}\sum_{\bf p_1p_2q}
 a_{{\bf p_1+q}\downarrow}^\dagger
 \left(a_{{\bf p_2-q}\uparrow}^\dagger a_{{\bf p_2}\uparrow}
 -\langle a_{{\bf p_2-q}\uparrow}^\dagger a_{{\bf p_2}\uparrow}
  \rangle\right)
 a_{{\bf p_1}\downarrow}.
\end{eqnarray}
Here $N$ is the number of lattice sites and 
$\langle\cdots\rangle$ denotes the average over the ground state.
Momentum ${\bf k}$ is defined in the first Brillouin zone, {\em i. e.},
$-\pi<k_x\leq \pi$ and $-\pi<k_y\leq \pi$ in units of $1/a$ with $a$
being the nearest neighbor distance for the square lattice.
The kinetic energy with momentum ${\bf k}$ is given by
$\epsilon_{\bf k}=-2t(\cos k_x + \cos k_y)$ for the square lattice.
The occupied electron density $n$ is defined by
\begin{equation}
 n\delta_{\bf q,0}=\frac{1}{N}\langle\sum_{\bf p}
  a_{{\bf p-q}\uparrow}^\dagger
   a_{{\bf p}\uparrow}\rangle.
\label{eq:ocu}
\end{equation}

The single-particle Green's function is defined by
\begin{equation}
 G_{\sigma}({\bf k},t)=-i\langle T[a_{{\bf k}\sigma}(t)
   a_{{\bf k}\sigma}^\dagger(0)]\rangle ,
\end{equation}
where $T$ represents the time ordering, and
$a_{{\bf k}\sigma}(t)={\rm exp}(i(H-\mu N_e)t)a_{{\bf k}\sigma}
{\rm exp}(-i(H-\mu N_e)t)$ with $\mu$ and $N_e$ being the chemical
potential and the number operator of electrons.
On the saturated ferromagnetic ground state, 
only the retarded part does not vanish with the down-spin electron, {\em i. e.},
\begin{equation}
 G_{\downarrow}({\bf Q},z)=-i\int_0^{\infty}
 \langle T[a_{{\bf Q}\downarrow}(t)a_{{\bf Q}\downarrow}^\dagger(0)]\rangle
 {\rm e}^{izt}{\rm d}t,\quad z=\omega+i\eta\quad (\eta\to 0^+).
\end{equation}
The unperturbed Green's function is given by
\begin{equation}
 G_{\downarrow}^{(0)}({\bf Q},z)^{-1}=z-\epsilon_{\bf Q}-Un+\mu,
\end{equation}
with $Un$ being the HF potential.
The self-energy is defined by 
\begin{equation}
 G_{\downarrow}({\bf Q},z)^{-1}=G_{\downarrow}^{(0)}({\bf Q},z)^{-1}
     -\Sigma_{\downarrow}({\bf Q},z).
\end{equation}

We sum up all the Feynman diagrams for the self-energy
shown in Fig.~\ref{fig.diag}, where only three lines of Green's function 
exist in the intermediate states. An important point is that
the frequency sum in the intermediate states can be explicitly carried out, 
since the Green's function for the down-spin electron consists only of 
a retarded part. After the frequency sum, we obtain
\begin{eqnarray}
 \Sigma_{\downarrow}({\bf Q},z)=
  \left(\frac{U}{N}\right)^2 &{\Bigl\{}&\sum_{\bf pq}G_0({\bf Q},z;{\bf pq}) 
   +\sum_{\bf pqq'}G_0({\bf Q},z;{\bf pq})\frac{U}{N}G_0({\bf Q},z;{\bf pq'}) \nonumber \\
  &+&\sum_{\bf pp'q}G_0({\bf Q},z;{\bf pq})\frac{-U}{N}
      G_0({\bf Q},z;{\bf p'q}) + \cdots {\Bigr\}},
\label{eq:diag}
\end{eqnarray}
where 
\begin{equation}
  G_0({\bf Q},z;{\bf pq}) = 1/(z-\epsilon_{\bf Q+p-q}-Un
     -\epsilon_{\bf q} +\epsilon_{\bf p}+\mu).
\end{equation}
Hereafter ${\bf p}$, ${\bf p'}$ are restricted inside the Fermi sphere, 
and ${\bf q}$, ${\bf q'}$ are restricted outside the Fermi sphere.

We notice that Eq.~(\ref{eq:diag}) is equivalent to collecting up all 
the multiple-scattering processes between two particles and one hole.
This three-body problem is exactly solvable by using 
the Faddeev equation.\cite{faddeev}
We write down the Faddeev equation for the scattering matrix $T(z)$;
\begin{equation}
  T(z) = T_1(z)+T_2(z),
\end{equation}
with
\begin{equation}
   \left(\begin{array}{c}
        T_1(z) \\
        T_2(z) 
\end{array}
    \right) =
      \left(\begin{array}{c}
        t_1(z) \\
        t_2(z) 
     \end{array}
     \right) +
    \left(\begin{array}{cc}
        0 & t_1(z) \\
        t_2(z) & 0 
     \end{array}
     \right) G_0(z)
     \left(\begin{array}{c}
        T_1(z) \\
        T_2(z) 
      \end{array}
     \right). \label{eq:fad}
\end{equation}
Here $G_0(z)$ is a free propagator $1/(z-H_0)$, $t_1(z)$ is 
the particle-particle scattering matrix, and $t_2(z)$ is 
the particle-hole scattering matrix.
All these quantities are represented within three-body states.
As shown in Fig.~\ref{fig.schema}, we introduce three-body states
with total momentum ${\bf Q}$,
which are written as
\begin{equation}
 |{\bf Q};{\bf pq}\rangle \equiv a_{{\bf Q+p-q}\downarrow}^\dagger
   a_{{\bf q}\uparrow}^\dagger a_{{\bf p}\uparrow}|F\rangle.
\label{eq:states}
\end{equation}
Here $|F\rangle$ is the saturated ferromagnetic ground state.
Note that $|{\bf Q};{\bf pq}\rangle$'s are orthonormal.
Then we have
\begin{eqnarray}
 \langle{\bf Q};{\bf pq}|\frac{1}{z-H_0}|{\bf Q};{\bf p'q'}\rangle &=&
 G_0({\bf Q},z;{\bf pq})\delta_{\bf p,p'}\delta_{\bf q,q'}, \\
 \langle {\bf Q};{\bf pq}|t_1(z)|{\bf Q};{\bf p'q'}\rangle
  & = & \frac{U}{N}\frac{1}{1-UD_1({\bf Q+p},z+\epsilon_{\bf p}-\mu)}
        \delta_{\bf p,p'},\\
  \langle {\bf Q};{\bf pq}|t_2(z)|{\bf Q};{\bf p'q'}\rangle
  & = & \frac{-U}{N}\frac{1}{1+UD_2({\bf Q-q},z-\epsilon_{\bf q}+\mu)}
        \delta_{\bf q,q'},
\end{eqnarray}
with
\begin{eqnarray}
 D_1({\bf k},z)  &=& \frac{1}{N}\sum_{\bf q'}
      \frac{1}{z-\epsilon_{\bf k-q'}-Un-\epsilon_{\bf q'}+2\mu},\\
 D_2({\bf k},z)  &=& \frac{1}{N}\sum_{\bf p'}
      \frac{1}{z-\epsilon_{\bf k+p'}-Un+\epsilon_{\bf p'}}.
\end{eqnarray}

For solving the Faddeev equation, we introduce the following quantities,
\begin{eqnarray}
 \Phi_{\rm pp}({\bf Q},z;{\bf p}) &=& \sum_{\bf p'q'}
     \langle {\bf Q};{\bf pq}|T_1(z)|{\bf Q};{\bf p'q'}\rangle 
       G_0({\bf Q},z;{\bf p'q'}), \\
 \Phi_{\rm ph}({\bf Q},z;{\bf q}) &=& \sum_{\bf p'q'}
     \langle {\bf Q};{\bf pq}|T_2(z)|{\bf Q};{\bf p'q'}\rangle 
       G_0({\bf Q},z;{\bf p'q'}). 
\end{eqnarray}
It will become clear below that $\Phi_{\rm pp}$ is independent of 
${\bf q}$ and $\Phi_{\rm ph}$ is independent of ${\bf p}$.
Then the Faddeev equation is rewritten as
\begin{eqnarray}
\Phi_{\rm pp}({\bf Q},z;{\bf p}) &=& 
\frac{UD_1({\bf Q+p},z+\epsilon_{\bf p}-\mu)}
{1-UD_1({\bf Q+p},z+\epsilon_{\bf p}-\mu)} \nonumber\\
 &+& \frac{U}{1-UD_1({\bf Q+p},z+\epsilon_{\bf p}-\mu)}
  \frac{1}{N}\sum_{\bf q'} G_0({\bf Q},z;{\bf pq'})
  \Phi_{\rm ph}({\bf Q},z;{\bf q'}), 
\label{eq:fad1}\\
  \Phi_{\rm ph}({\bf Q},z;{\bf q}) &=& 
\frac{-UD_2({\bf Q-q},z-\epsilon_{\bf q}+\mu)}
{1+UD_2({\bf Q-q},z-\epsilon_{\bf q}+\mu)} \nonumber\\
 &+& \frac{-U}{1+UD_2({\bf Q-q},z-\epsilon_{\bf q}+\mu)}
  \frac{1}{N}\sum_{\bf p'} G_0({\bf Q},z;{\bf p'q})
  \Phi_{\rm pp}({\bf Q},z;{\bf p'}).
\label{eq:fad2}
\end{eqnarray}
An equation containing only $\Phi_{\rm ph}({\bf Q},z;{\bf q})$ is obtained 
by substituting Eq.~(\ref{eq:fad1}) into Eq.~(\ref{eq:fad2}).
This integral equation can be accurately solved for systems 
of fairly large size, since its variable is only ${\bf q}$ 
for fixed ${\bf Q}$ and $z$.
Using the solution, we express the self-energy as 
\begin{equation}
 \Sigma_\downarrow({\bf Q},z) =\left(\frac{U}{N}\right)^2
    \sum_{\bf pq} G_0({\bf Q},z;{\bf pq})[1+\Phi_{\rm pp}({\bf Q},z;{\bf p})
  + \Phi_{\rm ph}({\bf Q},z;{\bf q})].
\label{eq:self}
\end{equation}

There appear many poles determined from the condition 
$1+UD_2({\bf Q-q},z-\epsilon_{\bf q}+\mu)=0$ in the low-energy region
for the self-energy. They constitute a continuous spectrum 
in the single-particle spectra in the $N\to\infty$ limit.
This condition is reduced
to $z=\omega_{\bf Q-q}+\epsilon_{\bf q}-\mu$ with $\omega_{\bf k}$ 
being the spin-wave energy in the random phase approximation (RPA). 
The lowest energy of the continuum spectra becomes zero at 
${\bf Q}={\bf k}_F$, since $\omega_0+\epsilon_{{\bf k}_F}-\mu=0$. 
For relatively small values of $|Q|$,  bound states appear below the 
continuum. They are identified as ``quasi-particles", and their 
energy $E_{\bf Q}$'s are determined from 
\begin{equation}
  G({\bf Q},E_{\bf Q})^{-1}=E_{\bf Q}
    -\left[\epsilon_{\bf Q}-\mu+Un+\Sigma_{\downarrow}
    ({\bf Q},E_{\bf Q})\right]=0.
\end{equation}

Now we comment on a widely used approximation
called ``ladder" approximation, which only considers the particle-particle
ladder and particle-hole ladder separately.
This is given by neglecting the second terms in Eqs.~(\ref{eq:fad1}) 
and (\ref{eq:fad2}). As a results, the self-energy is given by
\begin{equation}
 \Sigma_{\downarrow}({\bf Q},z)= U^2
 \left\{\frac{1}{N}\sum_{\bf q}
\frac{D_2({\bf Q-q},z-\epsilon_{\bf q}+\mu)}
{1+UD_2({\bf Q-q},z-\epsilon_{\bf q}+\mu)}
 +\frac{1}{N}\sum_{\bf p}
\frac{UD_1({\bf Q+p},z+\epsilon_{\bf p}-\mu)^2}
{1-UD_1({\bf Q+p},z+\epsilon_{\bf p}-\mu)}\right\}.
\label{eq:ph2}
\end{equation}
The first term comes from the particle-hole channel,
which is interpreted as the electron-magnon interaction with bare vertex $U$.
The second term comes from the particle-particle channel,
and its contribution is much smaller than the first one in the strong
coupling regime (except in the low density). As shown in the next section,
the self-energy in this approximation is larger in order of magnitude 
than the present theory.

\subsection{Limiting cases}

We discuss several limiting cases
for clarifying the physical meaning of the present theory.

\subsubsection{Low density limit ($n\to 0$)}

We expand the following quantities up to the first order of $n$:
\begin{eqnarray}
\frac{-UD_2({\bf Q-q},z-\epsilon_{\bf q}+\mu)}
{1+UD_2({\bf Q-q},z-\epsilon_{\bf q}+\mu)}
&\approx&
\frac{-Un}{z-\epsilon_{\bf Q-q}-\epsilon_{\bf q}+2\mu}, \\
\frac{-U}{1+UD_2({\bf Q-q},z-\epsilon_{\bf q}+\mu)}
&\approx&
-U+\frac{U^2n}{z-\epsilon_{\bf Q-q}-\epsilon_{\bf q}+2\mu}.
\end{eqnarray} 
Substituting these relations into Eq.~(\ref{eq:fad2}), we notice
that the first term is order $n$ and 
so is the second term, if $\Phi_{\rm pp}({\bf Q},z;{\bf p})$ is order 1.
Therefore we can neglect $\Phi_{\rm ph}({\bf Q},z;{\bf q})$ 
in the lowest order of $n$, and thereby we obtain the lowest-order form 
of $\Phi_{\rm pp}({\bf Q},z;{\bf p})$ as 
\begin{equation}
 \Phi_{\rm pp}({\bf Q},z;{\bf p}) =
 \frac{UD_1({\bf Q+p},z+\epsilon_{\bf p}-\mu)}
{1-UD_1({\bf Q+p},z+\epsilon_{\bf p}-\mu)}.
\end{equation}
Substituting this relation into Eq.~(\ref{eq:self}) 
and using the relation $(1/N)\sum_{\bf q}G_0({\bf Q},z;{\bf pq})
=D_1({\bf Q+p},z+\epsilon_{\bf p}-\mu)$, we obtain in the lowest order
\begin{eqnarray}
 Un+\Sigma_{\downarrow}({\bf Q},z) &\approx&\frac{1}{N}
 \sum_{\bf p}\frac{U}{1-UD_1({\bf Q+p},z+\epsilon_{\bf p}-\mu)}\nonumber\\
 &\approx& \frac{Un}{1-U\frac{1}{N}\sum_{\bf q'}
     \frac{1}{z-\epsilon_{\bf Q-q'}-\epsilon_{\bf q'}+2\mu}}.
\end{eqnarray}
Here the chemical potential $\mu$ is located at the bottom of the band, 
so that ${\bf q'}$ runs over the entire first Brillouin zone.
This is nothing but the low-density expression Kanamori derived several 
decades ago.\cite{kanamori}
Thus the present theory covers naturally the exact
low-density limit.

\subsubsection{Limit to the half-filling ($n\to 1$)}

We expand the following quantities up to the first order of $1-n$:
\begin{eqnarray}
\frac{UD_1({\bf Q+p},z+\epsilon_{\bf p}-\mu)}
{1-UD_1({\bf Q+p},z+\epsilon_{\bf p}-\mu)}
&\approx&
\frac{U(1-n)}{z-\epsilon_{{\bf Q+p-q}_0}+\epsilon_{\bf p}-U}, \\
\frac{U}{1-UD_1({\bf Q+p},z+\epsilon_{\bf p}-\mu)}
&\approx&
U+\frac{U(1-n)}{z-\epsilon_{{\bf Q+p-q}_0}+\epsilon_{\bf p}-U}.
\end{eqnarray} 
Here ${\bf q}_0$ stands for the momentum at the top of the valence band,
{\em i. e.}, ${\bf q}_0=(\pi,\pi)$ for the square lattice.
Substituting these relations into Eq.~(\ref{eq:fad1}), we notice that 
the first term becomes order $1-n$ and so is the second term, 
if $\Phi_{\rm ph}({\bf Q},z;{\bf q})$ is order 1. 
Therefore we can neglect $\Phi_{\rm pp}({\bf Q},z;{\bf p})$ 
in the lowest order of $1-n$, and thereby we obtain 
the lowest-order form of $\Phi_{\rm ph}({\bf Q},z;{\bf q})$ as 
\begin{equation}
 \Phi_{\rm ph}({\bf Q},z;{\bf q}) =
   \frac{-U\frac{1}{N}\sum_{\bf p'}\frac{1}{z-\epsilon_{{\bf Q+p'-q}_0}+\epsilon_{\bf p'}-U}}
     {1+U\frac{1}{N}\sum_{\bf p'}\frac{1}{z-\epsilon_{{\bf Q+p'-q}_0}+\epsilon_{\bf p'}-U}}.
\label{eq:high1}
\end{equation}
Substituting this relation into Eq.~(\ref{eq:self})
and using a lowest-order relation $G_0({\bf Q},z;{\bf pq})
\approx 1/[z-\epsilon_{{\bf Q+p-q}_0}+\epsilon_{\bf p}-U]$, we obtain
up to the first order of $1-n$
\begin{equation}
 Un+\Sigma_{\downarrow}({\bf Q},z) \approx U-
  \frac{U(1-n)}{1+U\frac{1}{N}\sum_{\bf p'}\frac{1}{z-\epsilon_{{\bf Q+p'-q}_0}+\epsilon_{\bf p'}-U}}.
\label{eq:ph1}
\end{equation}
Here ${\bf p'}$ runs over the entire first Brillouin zone.

Since available unoccupied states are limited in the up-spin band, 
the particle-particle multiple-scattering process gives rise to  merely 
a higher-order contribution, in contrast to the low-density limit. 
Also, the processes creating more particle-hole pairs give rise to 
higher-oreder contributions. Thus Eq.~(\ref{eq:ph1}) is exact up to
the first order of $1-n$ under the assumption of the saturated ferromagnetic 
ground state. 
But this expression is not so useful, since the saturated ferromagnetism is
always unstable in this limit by the reason discussed below.

The second term of Eq.~(\ref{eq:ph1}) vanishes as $n\to 1$, and 
one may think that the saturated ferromagnetic ground state is always
stable around $n \sim 1$. This is not true, since the second term of 
Eq.~(\ref{eq:ph1}) has a pole at $z=\omega_{{\bf Q-q}_0}$, which is negative:
$\Sigma_{\downarrow}({\bf Q},z)\propto 1/(z-\omega_{{\bf Q-q}_0})$
for $z\sim\omega_{{\bf Q-q}_0}$. Here $\omega_{\bf k}$ represents the
RPA spin-wave energy at $n=1$.
Therefore $G_\downarrow({\bf Q},z)^{-1}$ can be zero for 
$z\sim\omega_{{\bf Q-q}_0}$, even if $1-n$ is very small.
This means that the system is unstable through the coupling to 
the {\em unstable} spin-wave excitation.
Figure \ref{fig.magnon} shows $\omega_{\bf q}$ for several 
values of $U$ on the square lattice,
demonstrating that it is always negative.

\subsubsection{Atomic limit ($t\to 0$, $U<\infty$)}

In the limit of $t\to 0$, we have
\begin{equation}
  G_0({\bf Q},z;{\bf pq}) \sim \frac{1}{z-Un},\quad
  D_1({\bf q},z) \sim \frac{1-n}{z-Un},\quad
  D_2({\bf q},z) \sim \frac{n}{z-Un} .
\end{equation}
Substituting these relations into Eqs.~(\ref{eq:fad1}) and (\ref{eq:fad2}), 
we solve the Faddeev equation. The solution is given by
\begin{equation}
 \Phi_{\rm pp}({\bf Q},z;{\bf p}) \sim \frac{U(1-n)}{z-U(1-n)},\quad
 \Phi_{\rm ph}({\bf Q},z;{\bf q}) \sim \frac{-Un}{z-U(1-n)}.
\end{equation}
Substituting these relations into Eq.~(\ref{eq:self}),
we obtain the self-energy and the Green's function as
\begin{eqnarray}
 \Sigma_\downarrow({\bf Q},z)&=& U^2\frac{n(1-n)}{z-U(1-n)},\\
 G_\downarrow({\bf Q},z)&=& \frac{1-n}{z} + \frac{n}{z-U}.
\end{eqnarray}
This is the exact form in the atomic limit,
first emphasized by Hubbard.\cite{hubbard} 
The fact that the three-body scattering approximation
satisfies the exact atomic limit was first proved by one of the present 
author.\cite{iga1}

\subsubsection{Strong coupling limit ($U\to\infty$)}

We expand $\Phi_{\rm pp}({\bf Q},z;{\bf p})$ and 
$\Phi_{\rm ph}({\bf Q},z;{\bf q})$ in powers of $1/U$,
\begin{eqnarray}
 \Phi_{\rm pp}({\bf Q},z;{\bf p})&=&\Phi_{\rm pp}^{(0)}({\bf Q},z;{\bf p})
                                 +\Phi_{\rm pp}^{(1)}({\bf Q},z;{\bf p})+\cdots,
\label{eq:expand1}\\
 \Phi_{\rm ph}({\bf Q},z;{\bf q})&=&\Phi_{\rm ph}^{(0)}({\bf Q},z;{\bf q})
                                 +\Phi_{\rm ph}^{(1)}({\bf Q},z;{\bf q})+\cdots.
\label{eq:expand2}
\end{eqnarray}
Expanding also $D_1({\bf k},z)$ and $D_2({\bf k},z)$ in powers of $1/U$
and substituting them into Eqs.~(\ref{eq:fad1}) and (\ref{eq:fad2}),
we derive the equations for each order of $1/U$.
The derivation of the solutions
is straightforward but lengthy. We have summarized the derivation
in the Appendix A, and here we simply show the result:
\begin{eqnarray}
 \Phi_{\rm pp}^{(0)}({\bf Q},z;{\bf p}) &=& -1,\quad
 \frac{1}{N}\sum_{\bf p}\Phi_{\rm pp}^{(1)}({\bf Q},z;{\bf p})
   =-\frac{F({\bf Q},z)}{U}, 
 \label{eq:relat1}\\
 \Phi_{\rm ph}^{(0)}({\bf Q},z;{\bf q}) &=&
   \frac{F({\bf Q},z)}{J({\bf Q},z;{\bf q})},
 \quad \frac{1}{N}\sum_{\bf q}\Phi_{\rm ph}^{(1)}({\bf Q},z;{\bf q})
     =\frac{F({\bf Q},z)}{U},
\label{eq:relat2}
\end{eqnarray}
where
\begin{eqnarray}
 J({\bf Q},z;{\bf q})&=& z-\epsilon_{\bf q}+\mu-
                       \frac{1}{n}\frac{1}{N}\sum_{\bf p'}
     (\epsilon_{\bf Q+p'-q}-\epsilon_{\bf p'}),
\label{eq:j} \\
 F({\bf Q},z) &=& \frac{n}{\frac{1}{N}\sum_{\bf q}
  \frac{1}{J({\bf Q},z;{\bf q})}}.
\label{eq:relat3}
\end{eqnarray}
Substituting these solutions into Eq.~(\ref{eq:self}), 
and with the help of the expansion, $G_0({\bf Q},z;{\bf pq})\approx
-[1/(Un)+(z-\epsilon_{\bf Q+p-q}-\epsilon_{\bf q}
 +\epsilon_{\bf p}+\mu)/(Un)^2]$,
we obtain the self-energy
\begin{equation}
 \Sigma({\bf Q},z)= -Un-F({\bf Q},z).
\label{eq:suinfty}
\end{equation}
The first term cancels out the HF potential.
This expression is different from the previously proposed one
(Eq.~(32) in Ref.~8). We verify numerically in the next section that
the present expression gives the same quasi-particle energy 
as Roth's variational wave function.

\subsection{Relation to the variational principle}

We consider a system of $N_e$ up-spin electrons and one down-spin electron
with total momentum ${\bf Q}$.
Let the wave function be
\begin{equation}
 |\psi({\bf Q})\rangle =\alpha |{\bf Q}\rangle
     +\sum_{\bf pq}\beta_{\bf pq}|{\bf Q};{\bf pq}\rangle.
\label{eq:vari}
\end{equation}
Here $|{\bf Q}\rangle=a_{{\bf Q}\downarrow}^\dagger|F\rangle$, and 
$|{\bf Q};{\bf pq}\rangle$ was defined by Eq.~(\ref{eq:states}), where
${\bf p}$ and ${\bf q}$ are restricted inside and outside the Fermi sphere
in the up-spin band, respectively.
Since the expectation value of energy $E$ for this state
is given by $E=\langle\psi|H|\psi\rangle/\langle\psi|\psi\rangle$,
the extremum condition for $E$ is reduced to the following eigenvalue
equation:
\begin{eqnarray}
  H_{\bf Q;Q}\alpha + \sum_{\bf pq}H_{\bf Q;pq}\beta_{\bf pq}
    &=& E\alpha, \nonumber\\
  H_{\bf pq;Q}\alpha + \sum_{\bf p'q'}H_{\bf pq;p'q'}\beta_{\bf p'q'}
    &=& E\beta_{\bf pq}, 
\label{eq:eigen}
\end{eqnarray}
where $H_{\bf Q;Q}\equiv\langle {\bf Q}|H|{\bf Q}\rangle$, 
$H_{\bf Q;pq}\equiv\langle {\bf Q}|H|{\bf Q;pq}\rangle$,
$H_{\bf pq;Q}\equiv\langle{\bf Q;pq}|H|{\bf Q}\rangle$,
and $H_{\bf pq;p'q'}\equiv\langle{\bf Q; pq}|H|{\bf Q; p'q'}\rangle$.
Instead of solving directly this eigenvalue problem, we calculate
the resolvent $R(z)=1/(z-H)$ by using the relation $R(z)\cdot (z-H)=1$.
It is given by
\begin{equation}
 \langle{\bf Q}|R(z)|{\bf Q}\rangle
   =\left[ z-H_{\bf Q;Q}-\sum_{\bf pq;p'q'}H_{\bf Q;pq}
    S_{\bf pq;p'q'}(z)H_{\bf p'q';Q}\right]^{-1},
\label{eq:resolvent}
\end{equation}
where $S_{\bf pq;p'q'}(z)\equiv \langle{\bf Q; pq}|(z-H_3)^{-1}
|{\bf Q; p'q'}\rangle$ with $H_3$ being the Hamiltonian operator
represented by the components $\{H_{\bf pq;p'q'}\}$ 
within the three-body states.
The Faddeev equation is one way of calculating $S_{\bf pq;p'q'}(z)$.
Therefore, $\langle{\bf Q}|R(z)|{\bf Q}\rangle$ is essentially equivalent 
to $G({\bf Q},z)$ except for the origin of energy.
Note that the poles of $R(z)$ give the energy eigenvalues,
and let $E_i$'s be such energy eigenvalues. Then
the excitation energies of overturning an up-spin electron at the Fermi 
level and placing it on a down-spin band are given by
$E_i-\sum_{\bf p}\epsilon_{\bf p}-\mu$, which coincides with the
energies of the poles of $G({\bf Q},z)$. 

Now we compare the present theory with other variational theories.
The variational wave function used by Roth\cite{linden,roth} is given by
\begin{equation}
 |\Psi^R({\bf Q})\rangle=\sum_i{\rm e}^{i{\bf Qr}_i}a_{i\downarrow}^\dagger
 \left(1-\sum_j f(i-j)a_{j\uparrow}^\dagger a_{i\uparrow}\right)
 a_{{\bf k}_F\uparrow}|F\rangle,
\label{eq:roth}
\end{equation}
where ${\bf k}_F$ is a Fermi momentum, and $f(i-j)$ is the variational
parameter. The total momentum of this state is ${\bf Q-k}_F$.
The wave function used by Shastry {\em et al.}\cite{shastry} is given by
putting $f(i-j)=\delta_{i,j}$ in Eq.~(\ref{eq:roth}). 
Roth's wave function neglects
the correlation between the hole at the Fermi level (described 
by $a_{{\bf k}_F\uparrow}$) and other excitations.
If this hole at the Fermi level is disregarded, Roth's wave function
consists of three-body states of one particle-hole pair in the up-spin 
band and one electron in the down-spin band. Note that these states are 
limited in comparison with general three-body states, since two operators
are confined on the same site $i$ in Eq.~(\ref{eq:roth}).
This restriction and the condition $f(0)=1$ are necessary
for avoiding the double occupancy in the $U=\infty$ limit.
Therefore, we expect that the three-body scattering theory gives
the same results as Roth's in the $U=\infty$ limit.
Actually this is verified numerically in the next section.
Recently, Hanish {\em el al.}\cite{hanish} analyzed Roth's wave function 
on various lattices by using the resolvent method. They also
extended Roth's wave function for finite
$U$ by including more doubly occupied states such as
$a_{i\downarrow}^\dagger a_{i\uparrow}^\dagger a_{j\uparrow}$.
Their method has an advantage of being able to treat systems of infinite 
size, but their wave function is still limited in comparison with the general
three-body states, since two operators are still confined on the same site $i$.
On this point, the three-body scattering theory remains superior to
their approach. 
\section{Calculated Results}

Confining ourselves to finite systems of the square lattice, we calculate 
the single-particle Green's function. 
Since several $\epsilon_{\bf k}$'s are degenerate
for finite systems, shell structures appear on physical
quantities, depending on the occupation of up-spin electrons.
\cite{barbieri,linden}
The following numerical calculations are carried out on closed-shell 
configurations.
We first eliminate $\Phi_{\rm pp}({\bf Q},z;{\bf p})$ from
Eqs.~(\ref{eq:fad1}) and (\ref{eq:fad2}) to set up
the integral equation for $\Phi_{\rm ph}({\bf Q},z;{\bf q})$ only.
Then we solve the equation on systems of finite size;
the matrix of size $N(1-n)\times N(1-n)$ has to be inverted.

Figure \ref{fig.qp1} shows the quasi-particle energy and the lowest boundary
of the continuum spectra as a function of ${\bf Q}$, for
$\delta(\equiv 1-n)=0.2$ and $4t/U=0.05$.
The calculation was carried out on a system of $N=50\times 50$ with
a periodic boundary condition.
The dispersion of the quasi-particle energy $E_{\bf Q}$ is extremely flat 
in comparison with a free dispersion $\epsilon_{\bf Q}$. 
The effective mass $m^*$ is defined by
$E_{\bf Q}=E_0+(1/2m^*)Q^2$ around ${\bf Q}\sim 0$. The mass enhancement
$m^*/m$ ($m$ is the free mass given by $\epsilon_{\bf Q}$) reaches to 
as large as 12 for this situation.
The dispersion curve enters into the energy continuum for
$Q_x=Q_y > 0.4\pi$. The quai-particle peak may survive as a resonant peak
after entering the continuum spectra.
With increasing values of $|{\bf Q}|$, the boundary of the continuum spectra
goes down and eventually
touches to the $x$-axis at ${\bf Q}={\bf k}_F$.
A small gap seen at ${\bf Q}={\bf k}_F$ in the figure is due 
to a finite-size effect.

Figure \ref{fig.qp2}(a) shows the quasi-particle energy at ${\bf Q}=0$,
as a function of $4t/U$. The hole density and the system size are
the same as in Fig.~\ref{fig.qp1}, $\delta=0.2$ and $N=50\times 50$.
The $E_0$ decreases with decreasing values of $U$. It turns negative
for $4t/U>0.085$, indicating that the Nagaoka ground state 
becomes unstable by overturning an up-spin electron at the Fermi level 
and placing it on the bottom of the down-spin band.
Figure \ref{fig.qp2}(b) shows $E_0$ evaluated in the ladder approximation. 
It is found that the first term in Eq.~(\ref{eq:ph2}) has a dominant 
contribution.
The absolute values are larger in order of magnitude than the values 
of the three-body scattering theory, suggesting that the ladder
approximation overestimates the self-energy due to neglecting 
the vertex correction for the electron-magnon process. 
Note that the three-body scattering theory takes account of 
the vertex correction, since the {\em particle-hole}
multiple-scattering process is renormalized by the {\em particle-particle}
multiple-scattering process.

As already mentioned, the three-body scattering theory is expected
to give the same quasi-particle energy as Roth's variational wave
function in the $U=\infty$ limit.
We demonstrate this equivalence in Fig.~\ref{fig.inf}, which shows
$E_0$ as a function of $\delta$ in comparison with the result of Roth's 
wave function.\cite{linden,hanish} We have used Eq.~(\ref{eq:suinfty}) 
for the calculation on a system of size $N=100\times 100$.
Both results coincide with each other within the accuracy of numerical 
errors.

Now we discuss the phase diagram for the Nagaoka 
ground state. We determine the phase boundary from the condition 
$E_{\bf Q}<0$, which means the instability
with respect to overturning an up-spin electron at the Fermi level 
and placing it on the down-spin band.
Figure \ref{fig.sq} shows the phase diagram thus evaluated 
in comparison with the results of other theories.
Note that finite-size effects are not small even for relatively large size
$N=50\times 50$.
The solid line (SW) represents the boundary above which
the energy of spin waves becomes negative within the RPA.
The solid line  (RR) represents the boundary determined
by the method of Richmond and Rickayzen\cite{richmond} (see Appendix B). 
The broken line represents the boundary given by Hanish {\em et al.}
(the curve of ``RES3" in Ref.~24).\cite{com1}
All the curves have the similar slopes around $\delta\sim 0$, but depart
from each other with increasing values of $\delta$. The three-body
scattering theory gives the most severe condition for the phase
boundary, the smallest area of
the Nagaoka state. The critical hole concentration is given by 
$\delta_c=0.41$, the same as that of Hanish {\em et al.}
This is reasonable since both theories become equivalent to using 
Roth's variational wave function in the $U=\infty$-limit.

\section{Concluding Remarks}

We have developed a many-body theory which takes account of
the three-body multiple scattering, and have applied it to
the two-dimensional Hubbard model by assuming the saturated ferromagnetic 
ground state. We have solved numerically the Faddeev equation of
the three-body problem on systems of finite size, and have calculated
the single-particle Green's function. 
We have found that the energy scale of the quasi-particle is smaller
in order of magnitude than that of the ladder approximation.
This difference comes from the lack of the vertex correction to 
the electron-magnon scattering process in the ladder approximation.
Recently, a many-body theory called the fluctuation exchange approximation
(FLEX)\cite{bickers} is widely used in the paramagnetic phase.
This is a self-consistent version of the ladder approximation,
and has a nice feature of satisfying the conserving condition of Kadanoff
and Baym.\cite{kadanoff} One shortcoming of the FLEX is the lack
of the vertex correction. For this reason, it may be interesting 
to extend the three-body scattering theory to the paramagnetic phase by
setting up directly the integral equation instead of the Faddeev equation.

We have also studied the instability of the Nagaoka state.
The present theory is regarded as a natural extension of Roth's variational
wave function, having improved the phase diagram of the Nagaoka state.
This result may serve as a measure of the accuracy of the present theory. 

An accurate determination of the phase diagram is beyond the scope of 
this paper.
If one uses more sophisticated variational wave functions,
the area of the Nagaoka ground state can be reduced but remains finite.
One of the best phase boundary was obtained by Wurth {\em at al.},\cite{wurth}
who included more than 1000 terms with up to two particle-hole pair 
excitations in the variational calculation. Another interesting attempt 
was made by von der Linden and Edwards,\cite{linden}
who used the Slater determinant of the single-particle states of up-spin
electrons influenced by the moving down-spin electron. 
The latter approach is regarded as an extension of the method of Richmond 
and Rickayzen.\cite{richmond} Both gave smaller areas of the Nagaoka 
state than the present result. On the other hand,
Davis and Feldkamp\cite{davis} pointed out that the method of Richmond and 
Rickayzen can be extended to calculate accurately the whole spectra of the
single-particle excitations;
one first diagonalize all ground and excited states in the presence of
the static impurity of the down-spin electron and then diagonalize 
the hopping of the down-spin electron in this complete set of states.
Of course the actual calculation has to be approximate by 
truncating the number of the excited states. 
Davis and Feldkamp\cite{davis} and one of the present 
authors\cite{iga2} applied this method to finite-size systems of the 
one-dimensional Hubbard model, and calculated the whole spectral function.
This type of calculations has not been attempted yet
on the two-dimensional Hubbard model.

So far, we have considered only the single-particle Green's function.
The two-particle Green's function gives the poles of spin waves,
whose softening may set another condition for the instability of the 
Nagaoka state. 
The phase diagram obtained by Okabe\cite{okabe} along this line
is not better than the present one. 
As a natural extension of the present theory to calculate spin waves,
we need to consider four-body correlations.

Finally we comment on the application to realistic situations
of transition-metal compounds, in which 
five 3{\it d}-orbitals and the Hund-rule coupling are to be considered.
The present theory is too complicated to treat such situations,
and some simplifications have to be made.
We have simplified the calculational scheme by using 
the local approximation.
This type of calculations has been carried out for transition-metal 
mono-oxides\cite{taka1} and La$_2$CuO$_4$ and Sr$_2$CuO$_2$Cl$_2$,\cite{taka2}
which improves considerably the single-particle spectra of 
the band calculation based on the local density approximation
and leads to good agreement with the photoemission experiments.

\acknowledgments

We would like to thank Dr. T. Okabe for valuable discussion.
This work was partially supported by a Grant-in-Aid for Scientific Research
from the Ministry of Education, Science, Sports and Culture, Japan.

\appendix
\section{Derivation of Eqs.~(2.40) and (2.41)}

Expanding $D_1({\bf k},z)$ and $D_2({\bf k},z)$ in powers of $1/U$,
we rewrite Eqs.~(\ref{eq:fad1}) and (\ref{eq:fad2}) as
\begin{eqnarray}
 \Phi_{\rm pp}({\bf Q},z;{\bf p}) &=& -(1-n)-\frac{1-n}{U}
 (z+\epsilon_{\bf p}-\mu)+\frac{1}{U}A_{\rm pp}({\bf Q+p})+\cdots \nonumber\\
 &-&\frac{1}{N}\sum_{\bf q'}\left\{1+\frac{1}{U}(z+\epsilon_{\bf p}-\mu)
-\frac{1}{Un}[\epsilon_{\bf Q+p-q'}+\epsilon_{\bf q'}-2\mu
-A_{\rm pp}({\bf Q+p})]+\cdots \right\} \nonumber\\
  &\times&\Phi_{\rm ph}({\bf Q},z;{\bf q'}), \label{eq:strong1}\\
 \Phi_{\rm ph}({\bf Q},z;{\bf q}) &=& 
  \frac{-Un-(z-\epsilon_{\bf q}+\mu)+\frac{1}{n}A_{\rm ph}({\bf Q-q})}
       {J({\bf Q},z;{\bf q})}
 +\frac{\frac{1}{n}A_{\rm ph}'({\bf Q-q},z)}
       {J({\bf Q},z;{\bf q})^2}+\cdots
 \nonumber \\
 &+&\left\{\frac{-1}{J({\bf Q},z;{\bf q})}
          +\frac{\frac{1}{Un^2}A_{\rm ph}'({\bf Q-q},z)}
       {J({\bf Q},z;{\bf q})^2}+\cdots
    \right\} \nonumber \\
   &\times&\frac{1}{N}\sum_{\bf p'}\left\{
   U+\frac{z-\epsilon_{\bf Q+p'-q}-\epsilon_{\bf q}+\epsilon_{\bf p'}+\mu}{n} 
   +\cdots\right\}\Phi_{\rm pp}({\bf Q},z;{\bf p'}), \label{eq:strong2}
\end{eqnarray}
where
\begin{equation}
 J({\bf Q},z;{\bf q})=z-\epsilon_{\bf q}+\mu
 -\frac{1}{n}A_{\rm ph}({\bf Q-q}),
\end{equation} 
\begin{eqnarray}
 A_{\rm pp}({\bf k})&=&\frac{1}{N}\sum_{\bf q'} 
     (\epsilon_{\bf k-q'}+\epsilon_{\bf q'}-2\mu),\\
 A_{\rm ph}({\bf k})&=&\frac{1}{N}\sum_{\bf p'}
     (\epsilon_{\bf k+p'}-\epsilon_{\bf p'}), \\
 A_{\rm ph}'({\bf k},z)&=&\frac{1}{N}\sum_{\bf p'}
     (z-\epsilon_{\bf k+p'}+\epsilon_{\bf p'})^2 .
\end{eqnarray} 
Then we expand 
$\Phi_{\rm pp}({\bf Q},z;{\bf p})$ and 
$\Phi_{\rm ph}({\bf Q},z;{\bf q})$
in powers of $1/U$ as written down in Eqs.~(\ref{eq:expand1}) and 
(\ref{eq:expand2}). Substitute them into Eqs.~(\ref{eq:strong1}) and 
(\ref{eq:strong2}) and arranging the terms in each order of $1/U$, 
we obtain the following relations:
\begin{eqnarray}
 0 &=& -\frac{Un}{J({\bf Q},z;{\bf q})}
       -\frac{U}{J({\bf Q},z;{\bf q})}
        \frac{1}{N}\sum_{\bf p'}\Phi_{\rm pp}^{(0)}({\bf Q},z;{\bf p'}),
 \label{eq:asymp1}\\
\Phi_{\rm pp}^{(0)}({\bf Q},z;{\bf p}) &=& -(1-n)
       -\frac{1}{N}\sum_{\bf q'}\Phi_{\rm ph}^{(0)}({\bf Q},z;{\bf q'}),
 \label{eq:asymp2}\\
\Phi_{\rm ph}^{(0)}({\bf Q},z;{\bf q}) &=& -1
       +\frac{1}{J({\bf Q},z;{\bf q})^2}\frac{1}{n}A_{\rm ph}'({\bf Q-q},z)
 \nonumber\\
       &-&\frac{1}{nJ({\bf Q},z;{\bf q})}\frac{1}{N}\sum_{\bf p'}
        [z-\epsilon_{\bf Q+p'-q}-\epsilon_{\bf q}+\epsilon_{\bf p'}+\mu]
       \Phi_{\rm pp}^{(0)}({\bf Q},z;{\bf p'}) \nonumber\\
       &+&\frac{1}{n^2J({\bf Q},z;{\bf q})^2}
         A_{\rm ph}'({\bf Q-q},z)\frac{1}{N}\sum_{\bf p'}
       \Phi_{\rm pp}^{(0)}({\bf Q},z;{\bf p'}) \nonumber\\
       &-&U\frac{1}{J({\bf Q},z;{\bf q})}\frac{1}{N}\sum_{\bf p'}
       \Phi_{\rm pp}^{(1)}({\bf Q},z;{\bf p'}), 
 \label{eq:asymp3}\\
\Phi_{\rm pp}^{(1)}({\bf Q},z;{\bf p}) &=& -\frac{1-n}{U}
    (z+\epsilon_{\bf p}-\mu)+\frac{1}{U}A_{\rm pp}({\bf Q+p})\nonumber\\
       &-&\frac{1}{U}\frac{1}{N}\sum_{\bf q'}
      \left[ z+\epsilon_{\bf p}-\mu-\frac{1}{n}
      (\epsilon_{\bf Q+p-q'}+\epsilon_{\bf q'}-2\mu-A_{\rm pp}({\bf Q+p}))
      \right]\Phi_{\rm ph}^{(0)}({\bf Q},z;{\bf q'}) \nonumber \\
       &-&\frac{1}{N}\sum_{\bf q'}\Phi_{\rm ph}^{(1)}({\bf Q},z;{\bf q'}).
 \label{eq:asymp4} 
\end{eqnarray}
Note that ${\bf p,p'}$ are restricted inside the Fermi sphere
and ${\bf q,q'}$ are outside the Fermi sphere.

Equation (\ref{eq:asymp2}) indicates that
$\Phi_{\rm pp}^{(0)}({\bf Q},z;{\bf p})$ is independent
of ${\bf p}$. This fact and Eq.(\ref{eq:asymp1}) leads to
$\Phi_{\rm pp}^{(0)}({\bf Q},z;{\bf p})=-1$.
Substituting this relation into Eqs.~(\ref{eq:asymp2})
and (\ref{eq:asymp3}), we have
\begin{eqnarray}
 \frac{1}{N}\sum_{\bf q'}\Phi_{\rm ph}^{(0)}({\bf Q},z;{\bf q'}) &=&n, 
 \label{eq:asymp5}\\
 \Phi_{\rm ph}^{(0)}({\bf Q},z;{\bf q}) &=&
       -U\frac{1}{J({\bf Q},z;{\bf q})}\frac{1}{N}\sum_{\bf p'}
       \Phi_{\rm pp}^{(1)}({\bf Q},z;{\bf p'}).
\label{eq:asymp6}
\end{eqnarray}
Here other terms in Eq.~(\ref{eq:asymp3}) cancel out with each other. 
Equation (\ref{eq:asymp6}) indicates that the ${\bf q}$-dependence
of $\Phi_{\rm ph}^{(0)}({\bf Q},z;{\bf q})$ comes only from 
$J({\bf Q},z;{\bf q})$.
Therefore $\Phi_{\rm ph}^{(0)}({\bf Q},z;{\bf q})$ is written in a form
\begin{equation}
\Phi_{\rm ph}^{(0)}({\bf Q},z;{\bf q})=\frac{C}{J({\bf Q},z;{\bf q})}.
\label{eq:asymp7}
\end{equation}
The constant $C$ is determined by substituting Eq.~(\ref{eq:asymp7}) 
into Eq.~(\ref{eq:asymp5}):
\begin{equation}
 C\equiv F({\bf Q},z)= \frac{n}{\frac{1}{N}\sum_{\bf q}\frac{1}{J({\bf Q},z;{\bf q})}}.
\label{eq:asymp8}
\end{equation}
Substituting Eqs.~(\ref{eq:asymp7}) into Eq.~(\ref{eq:asymp6}), we obtain
\begin{equation}
 \frac{1}{N}\sum_{\bf p'}\Phi_{\rm pp}^{(1)}({\bf Q},z;{\bf p'})
  =-\frac{F({\bf Q},z)}{U}.
\label{eq:asymp9}
\end{equation}
Next, we rewrite Eq.~(\ref{eq:asymp4}) as
\begin{eqnarray}
 \Phi_{\rm pp}^{(1)}({\bf Q},z;{\bf p})
  &=& -\frac{1}{Un}\sum_{\bf q'}(z-\epsilon_{\bf Q+p-q'}
         -\epsilon_{\bf q'}+\epsilon_{\bf p}+\mu)
          \Phi_{\rm ph}^{(0)}({\bf Q},z;{\bf q'})\nonumber\\
   &-&\frac{1}{N}\sum_{\bf q'}\Phi_{\rm ph}^{(1)}({\bf Q},z;{\bf q'}),
\label{eq:asymp10}
\end{eqnarray}
with the help of Eq.~(\ref{eq:asymp5}).
Then we make sum with respect to ${\bf p}$ in both sides of 
Eq.~(\ref{eq:asymp10}). The first term on the right hand side
satisfies the relation 
\begin{equation}
 \frac{1}{N}\sum_{\bf p}(z-\epsilon_{\bf Q+p-q'}
         -\epsilon_{\bf q'}+\epsilon_{\bf p}+\mu)
          \Phi_{\rm ph}^{(0)}({\bf Q},z;{\bf q'})
    =nF({\bf Q},z),
\end{equation}
which is independent of ${\bf q'}$. Using this relation,
we have
\begin{equation}
 \frac{1}{N}\sum_{\bf p'}\Phi_{\rm pp}^{(1)}({\bf Q},z;{\bf p'})
  = -(1-n)\frac{F({\bf Q},z)}{U}
  -n\frac{1}{N}\sum_{\bf q'}\Phi_{\rm ph}^{(1)}({\bf Q},z;{\bf q'}).
\end{equation}
Substituting Eq.~(\ref{eq:asymp9}) into this equation, we finally obtain 
\begin{equation}
 \frac{1}{N}\sum_{\bf q'}\Phi_{\rm ph}^{(1)}({\bf Q},z;{\bf q'})
   =\frac{F({\bf Q},z)}{U}.
\end{equation}

\section{Adaption of the Method of Richmond and Rickayzen 
to Numerical Calculations}

Consider the situation that there exist $N_e$ up-spin electrons
and one down-spin electron. Neglecting the kinetic energy for the
down-spin electron, we assume that it sits at origin.
The up-spin electrons are under the static repulsive potential
localized at origin. Since there is no interaction working between 
up-spin electrons, we first calculate the single-particle energy by solving
the potential problem, and then make $N_e$ electrons occupy these
energy levels according to the Pauli principle.

Let the wave functions of single particles be
\begin{equation}
  |\psi^{(n)}\rangle = \sum_{\bf k}\alpha_{\bf k}^{(n)}|{\bf k}\rangle,
    \quad n=1,2,\cdots, N,
\end{equation}
where $|{\bf k}\rangle$ represents the normalized state of plane wave
with wave vector ${\bf k}$. Then the energy eigenvalue $E^{(n)}$ is 
given by
\begin{equation}
 \epsilon_{\bf k}\alpha_{\bf k}^{(n)}
  +\frac{U}{N}\sum_{\bf k'}\alpha_{\bf k'}^{(n)}
  = E^{(n)}\alpha_{\bf k}^{(n)}.
\end{equation}
We solve numerically this equation by directly diagonalizing 
the corresponding matrix for systems of finite size 
(up to $N=60\times 60$), although it is also possible to solve it
by using the Green's function. Lowest $N_e$ levels are
occupied by up-spin electrons, and thereby the
energy of the system is given by 
\begin{equation}
 E_{\rm tot}=\sum_{n=1}^{n=N_e}+ E_\downarrow,
\end{equation}
where $E_\downarrow=0$, since we neglected the kinetic
energy for down-spin electrons. The energy of the saturated
ferromagnetic ground state for $N_e+1$ up-spin electrons is given by
$E_g^0=\sum_{i=1}^{i=N_e+1}\epsilon_{{\bf k}_i}$.
If $E_{\rm tot}$ is smaller than $E_g^0$, the saturated
ferromagnetic ground state is unstable.

\begin{figure}

\caption{Feynman diagrams for the self-energy $\Sigma_\downarrow({\bf Q},z)$
within the three-body scattering theory. Only three lines of Green's 
functions exist in the intermediate states. The broken lines represent the
Coulomb interaction.
\label{fig.diag}}

\vskip 20pt

\caption{A sketch for a three-body state $|{\bf Q};{\bf pq}\rangle$.
\label{fig.schema}}

\vskip 20pt

\caption{Dispersion of spin-wave energy $\omega_{\bf q}$ 
along the line from ${\bf q}=(0,0)$ to $(\pi,\pi)$ within the RPA.
The density of up-spin electrons is $n=1$, and the system size is
$N=100\times 100$.
\label{fig.magnon}}

\vskip 20pt

\caption{Quasi-particle energy $E_{\bf Q}$ and the lowest boundary 
of the continuum spectra as a function of ${\bf Q}$. 
The calculation is carried out for a system of $N=50\times 50$.
The hole density and the Coulomb interaction are
$\delta(=1-n)=0.2$ and $4t/U=0.05$, respectively.
\label{fig.qp1}}

\vskip 20pt

\caption{Quasi-particle energy $E_{\bf Q}$ at ${\bf Q}=0$ as a function 
of $4t/U$. The hole density is $\delta=0.2$ and the system size is
$N=50\times 50$;
(a) the three-body scattering theory; (b) the ladder approximation.
The line with lower energy in (b) is the contribution only from the first 
term of Eq.~(\protect{\ref{eq:ph2}}).
\label{fig.qp2}}

\vskip 20pt

\caption{Quasi-particle energy $E_{\bf Q}$ at ${\bf Q}=0$ as a function of
$\delta$ in the $U=\infty$ limit. The system size is $N=100\times 100$. 
The solid line represents the result of Roth's wave function (taken from
Fig.~1 in Ref.~24).
\label{fig.inf}}

\vskip 20pt

\caption{Phase diagram for the square lattice. Open circles are for
a system of $N=40\times 40$, and the solid squares are for a system of
$N=50\times 50$. The broken line is the curve ``RES3" of Hanisch 
{\em et al.}\protect{\cite{hanish}}.
The solid line with letters ``SW" represents the boundary above which
the instability of the RPA spin-waves become unstable. 
The solid line with letters ``RR" represents
the boundary given by the method of Richmond and Rickayzen.
\label{fig.sq}}

\end{figure}

\begin{references}

\def\vol(#1,#2,#3){{\bf #1}, #3 (#2)}

\bibitem{hubbard}J. Hubbard, Proc. Royal Soc. London \vol(A276,1963,238).
\bibitem{kanamori}J. Kanamori, Prog. Theor. Phys. \vol(30,1963,275).
\bibitem{gutzwiller}M. C. Gutzwiller, Phys. Rev. Lett. \vol(10,1963,159).
\bibitem{hertz}J. A. Hertz and D. M. Edwards, J. Phys. \vol(F 3,1973,2174);
               D. M. Edwards and J. A. Hertz, J. Phys. \vol(F 3,1973,2191).
\bibitem{matsumoto}
H. Matsumoto, H. Umezawa, S. Seki, and M. Tachiki,
                   Phys. Rev. \vol(B 17,1978,2276).
\bibitem{iga1}J. Igarashi, J. Phys. Soc. Jpn. \vol(52,1983,2827).
\bibitem{iga2}J. Igarashi, J. Phys. Soc. Jpn. \vol(54,1985,260).
\bibitem{rucken}A. E. Ruckenstein and S. Schmitt-Rink,
 Int. J. Mod. Phys. \vol(B 3,1989,1809).
\bibitem{gor}The variational aspect of the three-body problem
has discussed by E. G. Goryachev, Phys. Lett. \vol(A 166,1992,148).
\bibitem{hsu}T. C. Hsu and B. Doucot, Phys. Rev. \vol(B 48,1993,2131).
\bibitem{nagaoka}Y. Nagaoka, Phys. Rev. \vol(147,1966,392).
\bibitem{thouless}D. J. Thouless, Proc. Phys. Soc. \vol(86,1965,893).
\bibitem{fang}Y. Fang, A. E. Ruckenstein, E. Dagotto and S. Schmitt-Rink,
 Phys. Rev. \vol(B 40,1989,7406).
\bibitem{doucot}B. Doucot and X. G. Wen, Phys. Rev. \vol(B 40,1989,2719).
\bibitem{barbieri}A. Barbieri, J. A. Riera, and A. P. Young,
                  Phys. Rev. \vol(B 41,1990,11697).
\bibitem{linden}W. von der Linden and D. M. Edwards, J. Phys.: Cond. Matter
\vol(3,1991,4917).
\bibitem{hirsch}J. Hirsch, Phys. Rev. \vol(B 31,1985,4403).
\bibitem{roth}L. M. Roth, Phys. Chem. Solids \vol(28,1967,1549);
Phys. Rev. \vol(186,1969,428).
\bibitem{allan}S. R. Allan and D. M. Edwards, J. Phys. \vol(F 12,1982,1203).
\bibitem{shastry}B. S. Shastry, H. R. Krishnamurthy and P. W. Anderson,
Phys. Rev. \vol(B 41,1990,2375).
\bibitem{basile}A. G. Basile and V. Elser, Phys. Rev. \vol(B 41,1990,4842).
\bibitem{okabe}T. Okabe, Prog. Theor. Phys. \vol(97,1997,21);
 Phys. Rev. \vol(B 57,1998,403).
\bibitem{richmond}P. Richmond and R. Rickayzen, J. Phys. \vol(C 2,1969,528).
\bibitem{hanish}T. Hanish, G. S. Uhlig and E. M\"uller-Hartmann, 
Phys. Rev. \vol(B 56,1997,13960).
\bibitem{faddeev}L. D. Faddeev, Zh. Eksperim. Teor. Fiz.
\vol(39,1960,1459) [English transl.:Soviet Phys.-JETP \vol(12,1961,1014)].
\bibitem{com1}This curve is very close to that of Roth's wave function.
See also the curve ``RES2" in Ref.~24.
\bibitem{bickers}N. E. Bickers and D. J. Scalapino,
Annals of Physics, \vol(193,1989,206).
\bibitem{kadanoff}L. P. Kadanoff and G. Baym,
{\em Quantum Statistical Mechanics} (Benjamin, Menlo Park, 1962).
\bibitem{wurth}P. Wurth, G. S. Uhrig and M\"uller-Hartmann,
Ann. Phys. (Leipzig) \vol(5,1996,148).
\bibitem{davis}L. C. Davis and L. A. Feldkamp, J. Appl. Phys.
\vol(50,1979,1944).
\bibitem{taka1}M. Takahashi and J. Igarashi, Ann. Phys. (Leipzig)
\vol(5,1996,247); Phys. Rev. \vol(B 54, 1996,13566);
Phys. Rev. \vol(B 56,1997,12818).
\bibitem{taka2}M. Takahashi and J. Igarashi, Phys. Rev. \vol(B 59,1999,) 
No.11.

\end{references}
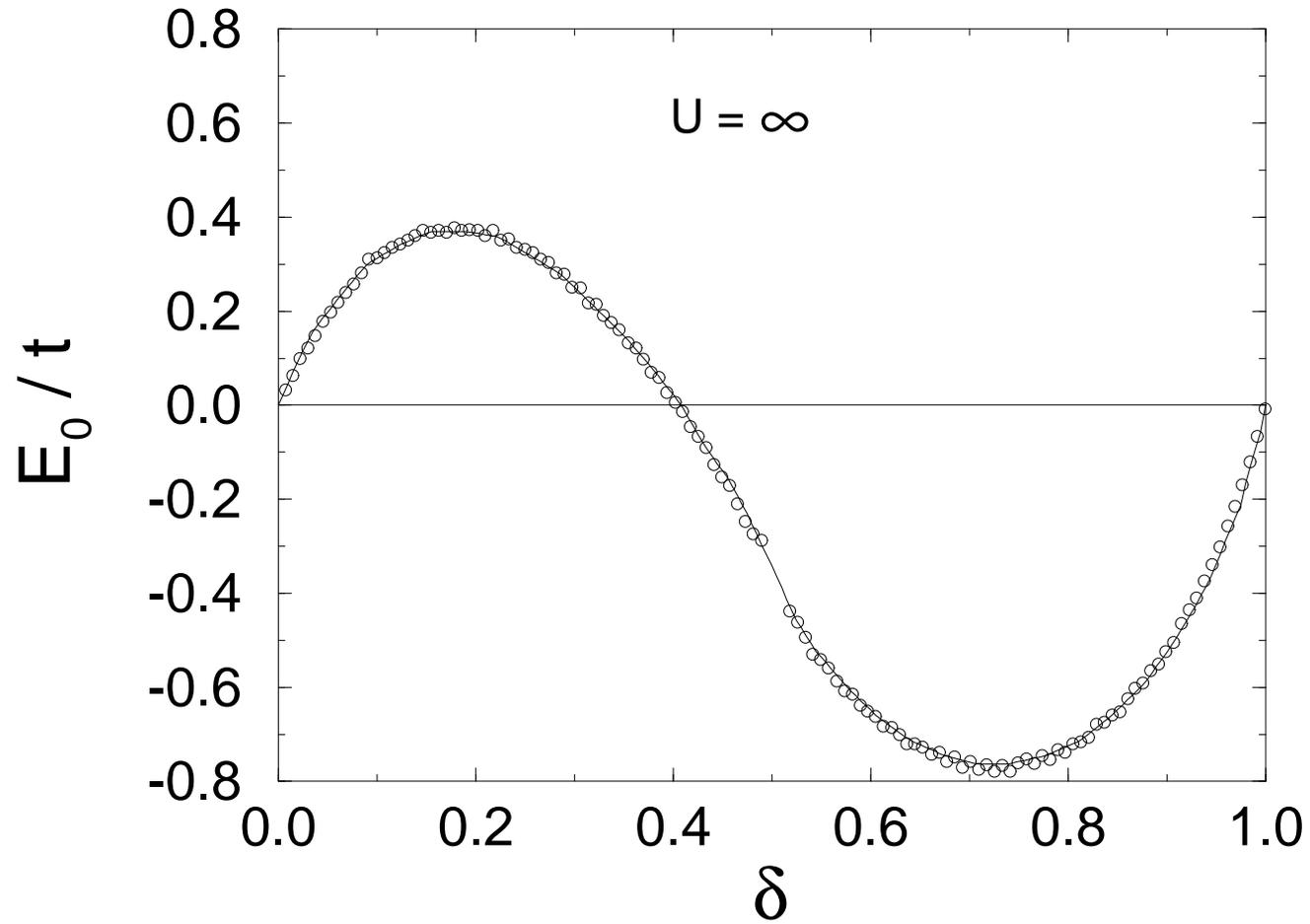
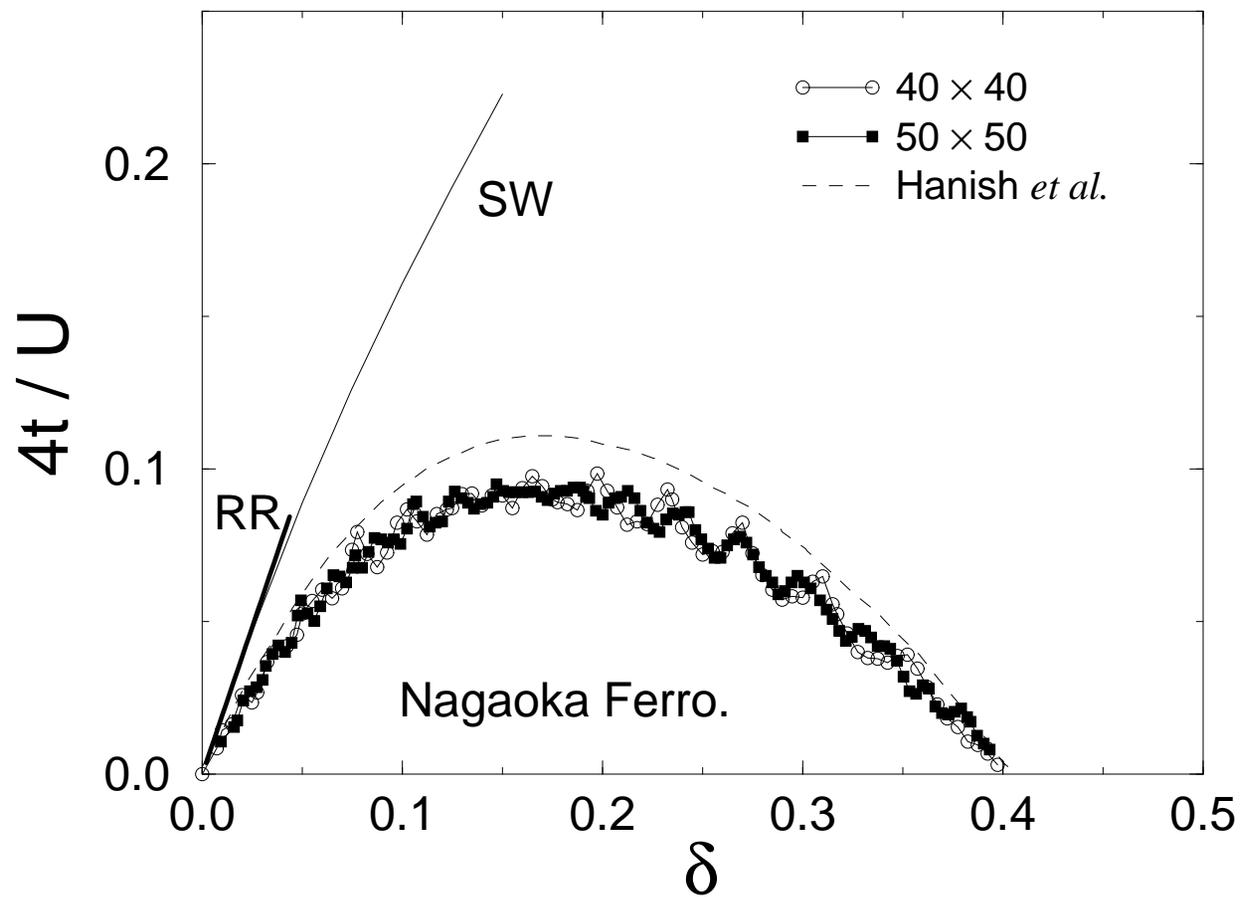
\end{document}